\documentclass[aps,prl,amsmath,amssymb,floatfix,reprint]{revtex4-2}

\usepackage{graphicx}
\usepackage{dcolumn}
\usepackage{bm}
\usepackage[usenames,dvipsnames]{xcolor}
\usepackage{siunitx}
\usepackage{braket}
\usepackage{amsmath}
\usepackage{mathptmx}
\usepackage{tcolorbox}
\usepackage{empheq}
\usepackage{physics}

\usepackage{multirow}
\usepackage{verbatim}
\usepackage{longtable}
\usepackage[normalem]{ulem}

\usepackage{float}

\begin{document}
	\title{On-demand source of dual-rail photon pairs based on chiral interaction in a nanophotonic waveguide}
	
	\author{Freja T. {\O}stfeldt}
	\author{Eva M. González-Ruiz}
	\author{Nils Hauff}
	\author{Ying Wang}
	\affiliation{Center for Hybrid Quantum Networks (Hy-Q), Niels Bohr Institute, University of Copenhagen, Blegdamsvej 17, DK-2100 Copenhagen, Denmark.}
	\author{Andreas D. Wieck}
	\author{Arne Ludwig}
	\author{Rüdiger Schott}
	\affiliation{Lehrstuhl f{\"u}r Angewandte Festk{\"o}rperphysik, Ruhr-Universit{\"a}t Bochum, Universit{\"a}tsstrasse 150, D-44780 Bochum, Germany.}
	\author{Leonardo Midolo}
	\author{Anders S. S\o{}rensen}
	\author{Ravitej Uppu}
	\author{Peter Lodahl}
	\email{lodahl@nbi.ku.dk}
	\affiliation{Center for Hybrid Quantum Networks (Hy-Q), Niels Bohr Institute, University of Copenhagen, Blegdamsvej 17, DK-2100 Copenhagen, Denmark.}
	
	\date{\today}
	\newpage
	
	\begin{abstract}
		Entanglement is the fuel of advanced quantum technology as it enables measurement-based quantum computing and allows loss-tolerant encoding of quantum information. In photonics, entanglement has traditionally been generated probabilistically, requiring massive multiplexing for scaling up to many photons. An alternative approach utilizing quantum emitters in nanophotonic devices can realize deterministic generation of entangled photons. However, such sources generate polarization-entanglement that is incompatible with spatial dual-rail qubit encoding employed in scalable photonic quantum computing platforms utilizing integrated circuits. Here we propose and experimentally realize an on-demand source of dual-rail photon pairs using a quantum dot in a planar nanophotonic waveguide. The source exploits the cascaded decay of a biexciton state and chiral light-matter coupling to achieve deterministic generation of spatial dual-rail Bell pairs with the amount of entanglement determined by the chirality. The operational principle can readily be extended to multi-photon entanglement generation required for efficient preparation of resource states for photonic quantum computing.
		
	\end{abstract}
	
	\maketitle
	
	\section{Introduction}
	Quantum photonics has developed significantly in the last decade and many of the required tools for scaling-up to advanced applications in quantum-information processing are already available \cite{OBrien2009,Uppu2021Perspective}.
	Indeed, the advent of production-mature advanced planar photonic-integrated circuits has been trumpeted as a major argument for photonic quantum computing\cite{Rudolph2017}. 
	Efficient photon sources, however, have remained challenging, and typically photon entanglement is heralded by multiplexing probabilistic sources\cite{Zhong2018}.
	An alternative approach exploits quantum emitters in photonic nanostructures that allow harvesting single-photon emission with near-unity efficiency\cite{Lodahl2015}.    
	This is the basis of deterministic photon sources that combined with optical switching can be demultiplexed to produce many simultaneous photons\cite{Lenzini2017,Wang2017}. 
	Quantum dot (QD) sources can generate two-photon polarization-entangled Bell states on-demand by exploiting the radiative decay of a biexciton state\cite{Benson2000,Akopian2006} to realize high entanglement fidelity and efficiency\cite{Huber2018,Liu2019}.
	These sources exploit polarization correlations induced from the biexciton radiative decay, which is not compatible with path encoding of photonic integrated circuits\cite{Politi2009}.
	Furthermore, rotationally-symmetric (in the plane perpendicular to the QD growth axis) photonic nanostructures are required to ensure equal coupling of orthogonal dipoles.
	This geometric constraint has restricted QD entanglement sources to the polarization degree-of-freedom, and most notably implied that the wide-bandwidth and highly-efficient planar nanophotonic waveguide platform \cite{Lodahl2015} is incompatible with biexciton entanglement generation. 
	We show that a chiral light-matter interface allows overcoming all of these limitations by realizing a spatial dual-rail encoded entanglement source.  
	
\begin{figure*}
		\centering
		\includegraphics[]{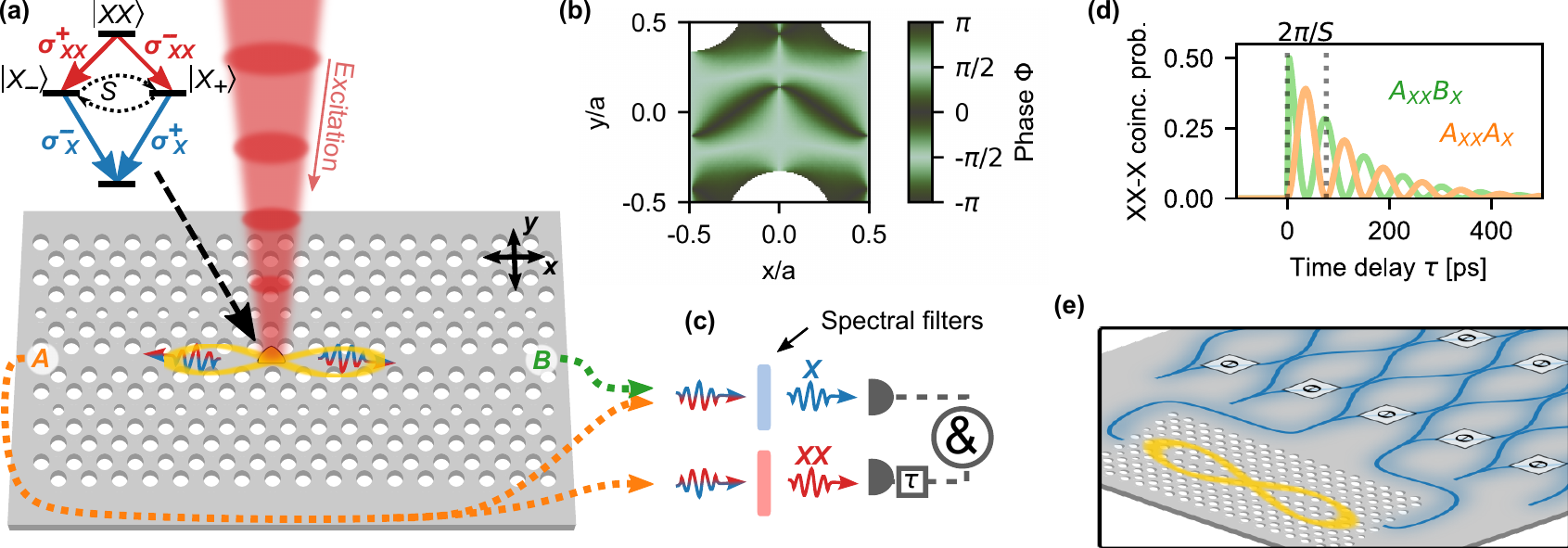}
		\caption{\label{fig:Fig1}
			\textbf{Operational principle of spatial dual-rail entanglement source in a planar photonic circuit.} 
			\textbf{(a)} Illustration of the device containing a QD in a glide-plane waveguide (GPW). The QD is optically excited to the biexciton state ($XX$) that recombines through a cascaded decay (level structure in inset) via the exciton states ($X_{\pm}$) emitting a circularly-polarized entangled photon pair. 
			In the GPW, circular-polarized photons with opposite helicity couples preferentially to only one of the directions, i.e. A or B, which enables path entanglement generation. A non-zero fine-structure splitting $S$ induces temporal spin flip oscillations between the exciton states, as illustrated on the inset.
			\textbf{(b)} Calculated local phase $\Phi$ between the $E_x$ and $E_y$ components of the electric field at different spatial locations within the unit-cell of the GPW. $\Phi = \pm \pi/2$ correspond to the locations of ideal chiral coupling. Here, $a$ is the lattice parameter of the photonic crystal. 
			\textbf{(c)} Experimental setup for projective exciton-biexciton cross-correlation measurements. 
			Two sets of cross-correlation measurements along the same path $(A_{XX}A_{X})$ and different paths $(A_{XX}B_{X})$ are performed. If the photons are collected from the same path, we split the signal using a 50:50 fiber beam splitter as illustrated with the orange dashed line.
			\textbf{(d)} Predicted two-photon cross-correlation dynamics (coincidence detection probability) for the experimental parameters of $S = 2\pi\times\SI{13}{\giga\hertz}$ and exciton radiative decay rate $\gamma_X = \SI{8.35}{\per\nano\second}$ and assuming ideal chiral coupling $\Phi = \pm \pi/2$. For details on the model see Supplementary material 5 \cite{Supplementary}.
			The experimental "tell tale" of path correlation is the phase offset of $2\Phi$ between the $(A_{XX}A_{X})$ and $(A_{XX}B_{X})$ correlation measurements. 
			\textbf{(e)} Proposed integration of a waveguide spatial dual-rail entanglement source into an advanced photonic integrated circuit with phase shifters ($\theta$) and beam splitters for quantum-information processing applications.}
	\end{figure*}

	The planar nanophotonic QD platform offers several salient features: it provides high-efficiency and broadband sources of highly indistinguishable photons\cite{Uppu2020} and is directly compatible with planar integrated photonic circuits for complex photonic processing\cite{Bogaerts2020}. 
	Integrated on-demand entanglement sources have however, been unattainable, since the high source efficiency relies on the radiative coupling of the QD to a single spatial and polarization mode. 
	This limitation can be overcome by exploiting {polarization-dependent} directional coupling \cite{Lodahl2017Chiral} where the photons from the QD are emitted preferentially into the left or the right propagating mode, labeled as $A$ and $B${, determined by the polarization $(\sigma^{\pm})$ of the optical transition,} cf. Fig. \ref{fig:Fig1}(a).
	{While glide-plane-symmetric photonic-crystal waveguides (GPWs)\cite{Sollner2015} and nanobeam waveguides \cite{Coles2016} have realized unidirectional coupling, residual backscattering of photons and multimode operation may adversely affect coherent quantum phenomena such as entanglement.
		In this article, we exploit specially-designed, single-mode GPWs with terminated mode adapters for minimized back scattering \cite{Mahmoodian2017,NilsSim} to transform the intrinsic polarization entanglement of the QD biexciton cascade into photon path entanglement.}
	Hereby dual-rail path encoded entanglement sources can be realized, which may be readily interfaced with photonic integrated circuits for realizing advanced quantum processors, cf. Fig. \ref{fig:Fig1}(e).
	
	\section{Path entanglement from chiral interactions}
	Figure \ref{fig:Fig1}(a) illustrates the operational principle of an on-demand source of path-entangled photons. 
	The QD biexciton state is excited through two-photon resonant excitation using optical laser pulses and decays through a correlated $\sigma^{\pm}$ - $\sigma^{\mp}$ cascade. 
	The GPWs are designed to feature simultaneously high directionality and coupling efficiency (i.e. $\beta$-factor) \cite{Sollner2015}, as required for an on-demand entanglement source. 
	For an ideally positioned QD $(\Phi = \pm \pi/2)$, $\sigma^+$- and $\sigma^-$-polarized transitions emit photons along opposite directions (denoted by $A$ and $B$) leading to optimal path entanglement. 
	The key governing parameter for the entanglement fidelity is the local phase $\Phi$ of the electric field at the position of the QD within the GPW with ideal performance at $\Phi=\pm \pi/2$ (see Supplementary material 5). Asymmetries in the  QD, however, introduce finestructure in the exciton level, which modifies the dynamics and leads to a photon state of the form\cite{bitheory}
	\begin{align}
	\begin{split}
	\ket{\psi(\tau)} = \frac{1}{\sqrt{N}}\Big( &\psi_{A_{XX}A_X}(\tau)\ket{A_{XX}A_{X}} + \psi_{A_{XX}B_X}(\tau)\ket{A_{XX}B_X} \\
	+ &\psi_{B_{XX}A_X}(\tau)\ket{B_{XX}A_{X}} + \psi_{B_{XX}B_X}(\tau)\ket{B_{XX}B_X} \Big) \,,
	\end{split}\label{eq:fullstate}
	\end{align}
	where $N$ is the normalisation factor and the amplitudes are 
	\begin{align}
	\begin{split}
	\psi_{A_{XX}A_{X}}(\tau) &=
	-\sqrt{2}\gamma_X e^{-\gamma_x \tau/2}\left( e^{-i(S \tau/2 +2\Phi)}+e^{iS \tau/2} \right) \\
	\psi_{B_{XX}B_X}(\tau) &=
	-\sqrt{2}\gamma_X e^{-\gamma_x \tau/2}\left( e^{-i(S \tau/2 -2\Phi)}+e^{iS \tau/2}\right) \\
	\psi_{A_{XX}B_X}(\tau) &= \psi_{B_{XX}A_X}(\tau) =
	-2\sqrt{2}\gamma_X\cos\left(S \tau/2\right)e^{-\gamma_X \tau/2}\,.
	\end{split}
	\label{eq:results1_approx}
	\end{align}
	Here, $\gamma_X$ is the exciton radiative decay rate, $S$ is the QD fine-structure splitting (in cyclic frequency units), and we assume that the radiative decay rate of the biexciton is $2\gamma_X$.
	We further assume that both the exciton and biexciton transitions couple to the waveguide with the same phase ($\Phi_X = \Phi_{XX}=\Phi$), which is a consequence of the wide operation bandwidth of GPWs compared to the energy difference between the exciton and biexciton states.
	For ease of notation, we omit the $X$ and $XX$ subscripts and adapt the convention that biexciton precedes the exciton, i.e., $AB$ represents $A_{XX}B_{X}$.
	In the case of a QD with $S\neq0$ positioned at a location with ideal chiral coupling $(\Phi=\pm\pi/2)$, a maximally-entangled path-encoded state is realized for all $\tau$, but oscillating between the two Bell states $\ket{\Psi}(\tau=0)=1/\sqrt{2}(\ket{AB}+\ket{BA})$ and $\ket{\Psi}(\tau=2\pi/S)=1/\sqrt{2}(\ket{AA}+\ket{BB})$. 
	The oscillation can be suppressed either by eliminating the fine-structure splitting $S$\cite{Huo2013} or by implementing time gating\cite{Huber2014}. 
	In path-resolved exciton($X$)-biexciton($XX$) photon correlations, cf. Fig. \ref{fig:Fig1}(c), the two-photon coincidence probabilities $|\bra{AA}\ket{\Psi(\tau)}|^2$ and $|\bra{AB}\ket{\Psi(\tau)}|^2$ are predicted to oscillate with a phase difference of $2 \Phi$, cf.  Fig. \ref{fig:Fig1}(d).
	In the absence of chiral coupling $(\Phi = 0)$ a separable state is generated, e.g.,  $\ket{\Psi}(\tau=0) = 1/2 (\ket{A_{XX}}+\ket{B_{XX}})(\ket{A_X}+\ket{B_X})$ and the phase shift in the correlation measurements is absent.
	The observation of a phase shift consequently constitutes the "tell tale" for the generation of path-encoded entanglement mediated by the chiral coupling.
	
	\section{Implementation}
\begin{figure*}
		\centering
		\includegraphics[]{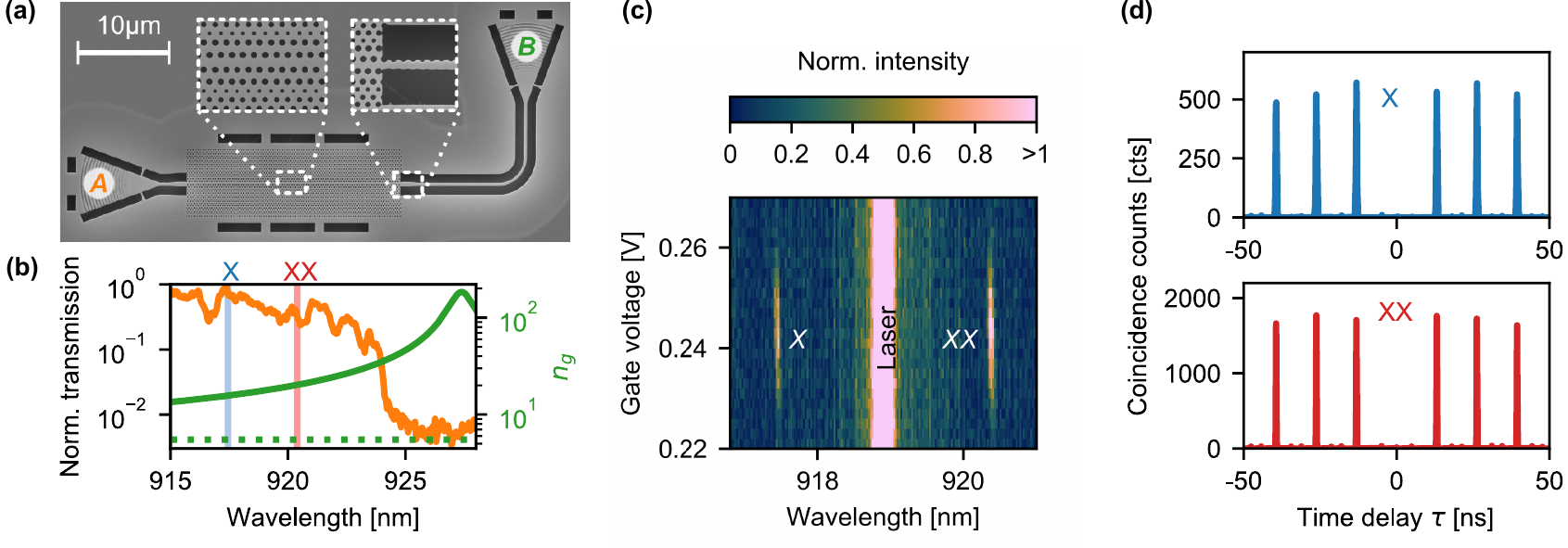}
		\caption{\label{fig:Fig2} 
			\textbf{Experimental characterization of QD coupled to a GPW.}
			\textbf{ (a)} Scanning electron microscope image of the device used for the measurements. The central section consists of a GPW (insets show close ups of the optimized waveguide design). 
			The GPWs are connected to grating outcouplers $A$ and $B$ via nanobeam waveguides. 
			The GPW-nanobeam interface features a mode adapter, as shown in the inset. 
			\textbf{(b)} Spectrally-resolved laser transmission shown in orange from gratings $A$ to $B$ for identifying the slow-light region. The curve is normalized to the transmission through a nanobeam waveguide (cf. Supplementary material 1). The green curves are the simulated group index $n_g$ for the GPW (solid line) and a nanobeam waveguide (dotted line).
			The wavelengths of the $X$ and $XX$ transitions of the QD are marked as well. 
			\textbf{(c)} Gate-voltage-dependent resonance fluorescence of the QD under two-photon pulsed excitation displaying the $X$ and $XX$ transitions. 
			\textbf{(d)} Pulsed second-order correlation measurements of $X$ and $XX$ transitions after spectral filtering of each transition ($\SI{22}{\giga\hertz}$ bandwidth).
			We extract $g_{X/XX}^{(2)}(0) \leq 0.009$ for both transitions indicating excellent single-photon purity.
		}
	\end{figure*}

	The experimental demonstration is realized with single QDs located in electrically-contacted waveguide samples for low-noise performance\cite{Pedersen2020} and Stark tuning of the charge states.
	We employ self-assembled indium arsenide (InAs) QDs epitaxially grown at the center of a suspended $~\SI{170}{\nano\meter}$ gallium arsenide (GaAs) membrane comprising a \textit{p-i-n} diode for charge control of the QD, c.f., Supplementary material 1 for details about the heterostructure. 
	The QDs are located randomly across the sample with an average density of $\approx \SI{1}{\per\micro\meter\squared}$. Figure \ref{fig:Fig2}(a) shows a scanning electron microscope image of the nanofabricated device in the GaAs membrane. 
	The device has a total footprint of $50 \times \SI{25}{\micro\meter\squared}$. 
	The central part of the device contains the GPW where the studied QD is located. 
	The GPW is designed with a lattice constant of $a = \SI{260}{\nano\meter}$ and hole radius of $r=\SI{69}{\nano\meter}$. 
	The glide plane geometry is made by shifting the holes by half a lattice constant ($a/2$) compared to a regular photonic-crystal waveguide along the propagation direction ($x$). Additionally, the geometry is optimized for single-mode operation, following the proposal of  Ref. \cite{Mahmoodian2017}, by shifting row $2-4$ of holes outwards (along $y$) by $0.25a\sqrt{3/2}$, $0.2a\sqrt{3/2}$, and $0.1a\sqrt{3/2}$, respectively, while the radii of holes in rows $1-3$ are also modified according to $r_{1,2} = 1.17r$ and $r_3 = 0.8r$. 
	A detailed discussion of the geometry is provided in Supplementary Information Fig. S2(a).
	The mode adapters \cite{Mahmoodian2017} (see inset of Fig. \ref{fig:Fig2}(a)) are designed to minimize reflections at the nanobeam/GPW interface that occurs due to a large group-index mismatch
	The nanobeam waveguides are terminated with two shallow-etched grating outcouplers \cite{zhou2018} that are oriented by $\SI{90}{\degree}$ with respect to each other for orthogonally polarized free-space collection.
	
	


	The sample is cooled to $\SI{1.6}{\kelvin}$ in a helium closed-cycle cryostat, with electrical and optical access. 
	The excitation laser is focused to the sample using a high numerical aperture  (NA$=0.81$) microscope objective. 
	A single QD is excited by a pulsed Ti:Sapphire laser ($\SI{20}{\pico\second}$ optical duration) whose frequency is tuned to satisfy the two-photon resonance condition of the QD biexciton.
	The emitted photons coupled to the GPW are directed out-of-plane at two high-efficiency grating outcouplers ($A$ and $B$ in Fig. \ref{fig:Fig2}(a)), and are collected by the same objective lens, and separated from the input excitation path using a $5:95$ beam splitter. 
	As the light diffracted by the outcouplers is linearly-polarized, the $\SI{90}{\degree}$ relative orientation of the outcouplers ensures cross-polarization in collection.
	We utilize this property to collect path-dependent emissions at the outcouplers $A$ and $B$ without compromising the collection efficiency.
	Using a set of quarter and half waveplates together with a polarizing beam splitter, we separate the emission from the two orthogonally polarized grating outcouplers into two separate spatial modes and couple each into a single-mode optical fiber.
	The emission at each of the outcouplers is spectrally filtered (bandwidth = $\SI{22}{\giga\hertz}$) and detected using a superconducting nanowire single-photon detector (SNSPD) with a low timing jitter $<\SI{15}{\pico\second}$. A time tagging device is used to register the timestamps of the detection events in both paths with a resolution of $\SI{4}{\pico\second}$.
	The timestamps accumulated over an hour are processed using MATLAB to construct the photon correlation histograms shown in Fig. \ref{fig:Fig3}.
	The low timing jitter of the SNSPDs is crucial in resolving the rapid oscillations (period $\approx\SI{80}{\pico\second}$) in the photon correlations induced by the fine-structure splitting of the QD excitons. 
	
	\section{Results}
	
	\subsection{QD characterization}
	The GPW device is characterized by spectrally-resolved transmission measurements where a tunable diode laser beam is injected into the waveguide at port $A$ and the transmitted light is measured at port $B$, cf. Fig. \ref{fig:Fig2}(b). 
	From the optical excitation of the QD we identify the spectral distance of the $X$ and $XX$ emission from the high group index ($n_g$) region. 
	The GPW experiences slow light with slow-down factors of $16$ and $14$ for the $XX$ and $X$ transitions, respectively. 
	This gives rise to slow-light induced Purcell enhancement and a near-unity coupling efficiency quantified through the $\beta$-factor. 
	We measure a radiative decay rate of $\gamma_X = \SI{8.35}{\per\nano\second}$ corresponding to a Purcell factor of about $2$ leading to $\beta > 95 \%$, i.e. near-deterministic operation of the source\cite{Sollner2015,Mahmoodian2017}, see  Supplementary material 3 for the experimental data. 
	Spectrally-resolved emission is collected at the out-coupling grating $B$ as a function of applied gate voltage, see Fig. \ref{fig:Fig2}(c), whereby $X$ and $XX$ transitions can be identified from their identical charge tuning slopes.
	Distinct anti-bunching is observed in pulsed second-order correlation measurements ($g^{(2)}(\tau)$) of both the $X$ and $XX$, cf. data in Fig. \ref{fig:Fig2}(d). 
	This confirms the high-purity single-photon emission from each transition, which is a prerequisite for entangled photon pair generation.
	
	\subsection{Path-correlation measurements}
	We quantify the generation of path-entangled photon pairs from path- and spectrally-resolved correlation measurements, with the resulting data displayed in Fig. \ref{fig:Fig3}(a). 
	The experimental data are very well explained by the theoretical model, cf. Supplementary material 5. 
	The correlation function decays with the independently measured exciton decay rate $\gamma_X$ (see Supplementary material 3), while the fine-structure splitting $S$ and phase shift $\Phi$ are extracted from the analysis.
	A key aspect to extract the phase, is high-precision time-matching of photon-correlation datasets, cf. Supplementary material 4.
	A pronounced phase shift of $\Phi = (0.12\pm0.01)\pi$ is extracted, which is the experimental evidence of chiral coupling resulting in path entanglement. 
	For comparison, Fig. \ref{fig:Fig3}(b) displays the separable case of a QD in a regular photonic-crystal waveguide, where we observe no pronounced phase shift. In this case the local electric field polarization is preferentially linear, meaning that pronounced chiral coupling is not expected. 
	
	\section{Observation of Path Entanglement}
	As an interesting note, the experiment directly probes the phase of the local electric field inside a complex photonic nanostructure. The measurements were repeated on two additional QDs (cf. Supplementary material 6) and $\Phi\approx0.1\pi$ was recorded in all cases. In a GPW, QDs are with hight probability positioned near points with $\Phi \approx \pi/2$, cf. Fig. \ref{fig:Fig1}(b). The experimental values are likely influenced by the presence of weak residual back-reflections from the outcoupling gratings, which may be overcome by further optimization of the grating design\cite{zhou2018}. Such weak scattering reduces the directionality, since a fraction of the emitted photons is back-reflected\cite{Lodahl2017Chiral}. See Supplementary material 6 for discussion and estimation of this effect. 
	
\begin{figure*}  
		\centering
		\includegraphics[]{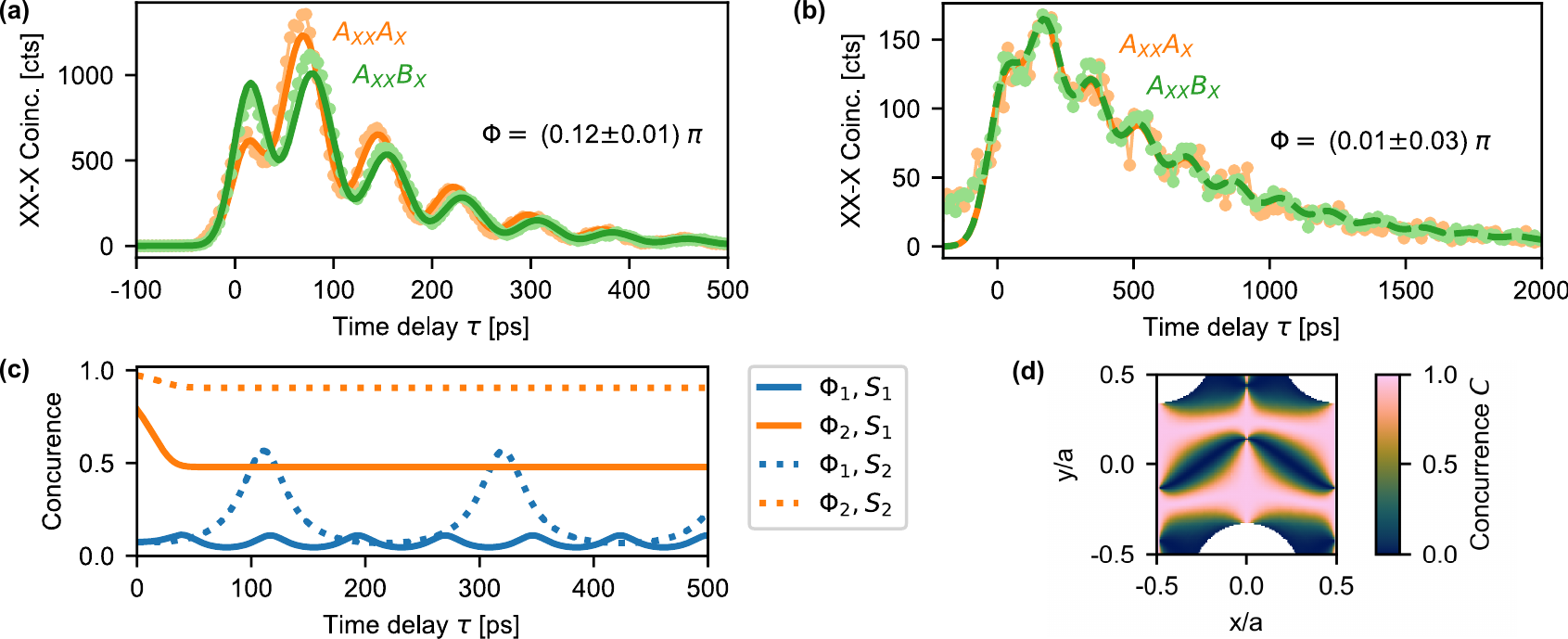}
		\caption{\label{fig:Fig3}
			\textbf{Observation of path-dependent correlations of the entangled state.} 
			\textbf{(a)} Path-dependent photon correlation measurements of QDs in GPW in two configurations, $A_{XX}B_{X}$ and $A_{XX}A_{X}$. Solid lines are fits to the theoretical model. The error in $\Phi$ is the $1\sigma$ confidence interval fitting error.
			\textbf{ (b)} Similar measurements as (a) but from a QD in a regular photonic crystal waveguide with no significant chiral interaction. 
			\textbf{(c)} Time-dependent concurrence $C$ calculated for a QD in the GPW under current experimental conditions ($\Phi_1 = 0.12\pi$, $S_1 = 2\pi\times \SI{12.78}{\giga\hertz}$, $\varsigma = \SI{15}{\pico\second}$), blue solid curve.
			The calculation for  $\Phi_2 = \pi/2$ and otherwise same parameters is shown as the orange solid curve.
			The dashed curves correspond to the case of a QD with lower fine-structure splitting of $S_2 = 2\pi\times \SI{5}{\giga\hertz}$.
			\textbf{(d)} Predicted QD position-dependent $C$ of the path-entangled state at time $\tau=0$ within the GPW exhibiting wide areas of near-perfect entanglement. The concurrence is calculated using the theoretical model\cite{bitheory}, and the simulated phase from Fig. \ref{fig:Fig1}(b).
		}
	\end{figure*}

	The local phase $\Phi$ is a governing parameter determining the quality of the generated path entanglement. 
	The degree of path entanglement is quantified by the concurrence $C$, where $C > 0$ is an entanglement criterion.
	Furthermore, the interplay between the emission-time uncertainty, fine-structure splitting induced oscillations, and timing jitter $\varsigma$ of the detectors determines the observable concurrence, as described by the theoretical model of the biexciton cascaded decay in a chiral waveguide \cite{bitheory}.
	Figure \ref{fig:Fig3}(c) plots the time evolution of the concurrence for real experimental parameters investigating the role of chirality $\Phi$ and fine-structure splitting $S$.
	The initial decay is determined by the interplay between the emission time and detector jitter, through the ratio $\gamma_X/\varsigma$, where a smaller ratio leads to a lower entanglement degradation.
	$C$ oscillates with a period of $2\pi/S$ reaching up to $0.11$ for our current experiment, a testimony of the successful generation of path entanglement by the source.
	$C$ can be dramatically increased by lowering the ratio of $S/\varsigma$. 
	With the fine-structure splitting of $S = 2\pi \times \SI{5}{\giga\hertz}$ recently reported on a similar device platform\cite{Uppu2020}, we estimate $C\approx0.57$ without any other experimental improvements ($\Phi$ and $\varsigma$). 
	{Further, droplet-etched GaAs QDs are known to possess $S \approx 0$, which could be ideal candidates for on dual-rail entanglement\cite{Huber2018}.}
	Increasing $\Phi$ improves the entanglement quality significantly and $C > 0.9$ is readily attainable for $\Phi = \pi/2$, cf. Fig. \ref{fig:Fig3}(c).
	The QD position-dependence of the maximum concurrence realizable in a GPW is displayed in Fig. \ref{fig:Fig3}(d), i.e. the GPW supports path entanglement generation with near-unity fidelity within a wide area of the waveguide.
	
	In summary, we have proposed and experimentally implemented an on-demand planar nanophotonic waveguide source of dual-rail path-entangled photon pairs. 
	The operation of the source is based on chiral coupling of a QD to the waveguide mode leading to directional emission. 
	The operational principle is general, and could readily be extended also to multi-photon entanglement sources, e.g., for the generation of path-encoded photonic cluster states\cite{Lindner2009}. 
	Such deterministic entanglement sources are much needed for advanced quantum-information processing applications \cite{Walther2005,Azuma2015}. 
	For instance, in photonic quantum computing, the most demanding task is the generation of high-quality multi-photon entangled resource states\cite{bartolucci2021} with photonic integrated circuits providing a realistic route for realizing quantum processors\cite{elshaari2020natphot}. 
	Consequently, interfacing QD path-entanglement sources with mature integrated photonic circuits, as sketched in Fig. \ref{fig:Fig1}(e), appears a resource-efficient approach. 
	To this end, heterogeneous chip transfer processes may be applied in order to combine the active QD devices and the passive photonic circuits\cite{wan2020nature}.

	
	\section*{Data availability}
	The data presented in the figures of this study are available from the authors on request.
	
	\section*{Acknowledgements}
	We thank Single Quantum for giving us access to the ultra-low timing-jitter superconducting nanowire single-photon detectors required for extracting the correlations. We gratefully acknowledge financial support from Danmarks Grundforskningsfond (DNRF 139, Hy-Q Center for Hybrid Quantum Networks). 
	This project has received funding from the European Union's Horizon 2020 research and innovation programmes under grant agreement No. 824140 (TOCHA, H2020-FETPROACT-01-2018) and under the Marie-Sk\l{}odowska-Curie grant agreement No. 801199.
	A.D.W. and A.L. acknowledge gratefully support of DFG-TRR160,  BMBF - Q.Link.X  16KIS0867, and the DFH/UFA  CDFA-05-06.
	We acknowledge Adam Knorr for initial sample characterization. We thank Matthew Broome for discussions in the early phase of the project.

	\newpage
	\onecolumngrid 
		
	\appendix
	
	\section*{SUPPLEMENTARY MATERIAL}
	\subsection{Sample heterostructure layout and fabrication}\label{sec:SI:Fab}
\begin{figure}
		\centering
		\includegraphics[]{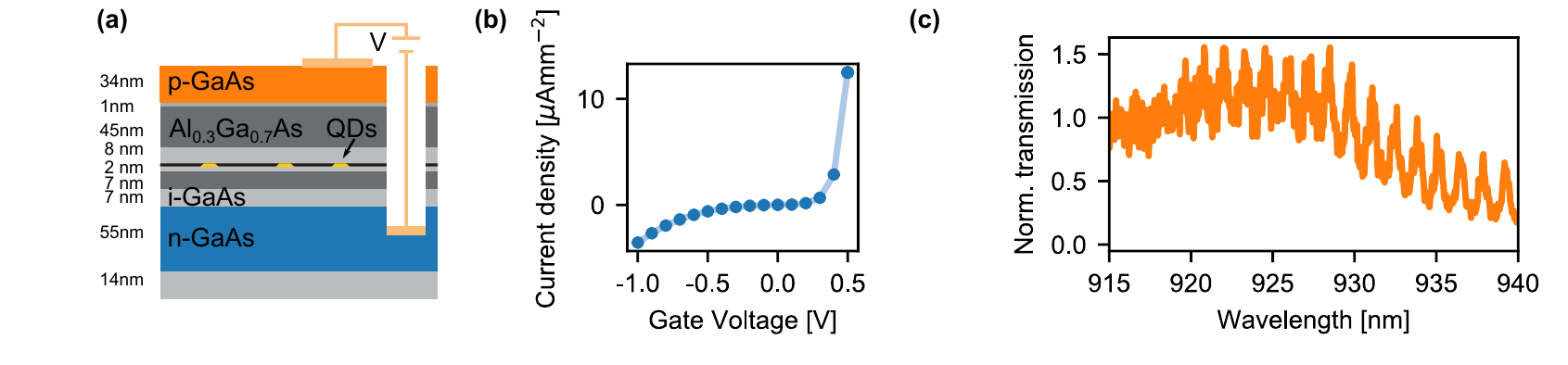}
		\caption{\label{fig:IV}\textbf{Electrical sample properties}. \textbf{(a)} Layout of the membrane containing \textit{p-i-n} doped layers for electrical control, to which external electrical contacts are bonded. The QDs are located in the center of the membrane.  \textbf{(b)} Current-voltage (IV) curve across the diode, showing small amounts of leakage current. The current density is found by dividing the current by the mesa size which is $2\times\SI{2}{\milli\meter\squared}$. \textbf{(c)} Transmission of a narrow band tunable laser through a nanobeam waveguide. The oscillations comes from back reflections from the grating couplers. The transmission scan is used for nomalisation of the GPW transmission scan in Fig. 2(b).}
	\end{figure}

	The sample membrane is grown using molecular beam epitaxy on a (100) GaAs substrate. 
	First a sacrificial layer with Al$_{0.75}$Ga$_{0.25}$As of $\SI{1371}{\nano\meter}$ thickness is grown to suspend the GaAs membrane after wet etching. 
	A layout of the membrane heterostructure, is displayed in Fig. \ref{fig:IV}(a), with the QDs in the center. 
	P-type Carbon and n-type Silicon doped layers in the membrane form an ultra thin \textit{p-i-n} diode around the QDs for charge state control and Stark tuning of the QD, which simultaneously suppresses charge noise. We use Carbon as p-type doping due to its high doping efficiency and very low diffusivity.
	The Al$_{0.30}$Ga$_{0.7}$As layers above and below the QD allow charging the QD near $\SI{0}{\volt}$ and prevent high currents to flow when the diode is used in forward bias voltage. 
	Electrical contacts are deposited on the \textit{n} and \textit{p} layers and trenched to create a single mesa of $2$ mm $\times~2$ mm to which the external voltage is applied, as also illustrated in Fig. \ref{fig:IV}(a). 
	Using reactive ion etching (RIE) vias are opened to the buried n-layer and Ni/Ge/Au/Ni/Au contacts are deposited by electron-beam physical vapor deposition and annealed at $\SI{430}{\celsius}$.
	Cr/Au pads are deposited on the p-layer to form Ohmic contacts.
	The nanophotonic structures are fabricated with electron-beam lithography at $\SI{100}{\kilo\electronvolt}$ (Elionix ELS 7000) and etched in the GaAs membrane using inductively-coupled plasma RIE in a BCl$_3$/Cl$_2$/Ar chemistry. 
	The undercut and sample cleaning was performed using the procedure discussed in Ref. \onlinecite{Midolo2015}.
	
	The contacts are connected to wire bonding pads, to which the external electrical connections are bonded. 
	The quality of the diode structure and the electrical contacts is checked by recording a current-voltage (IV) curve of the sample as seen in Fig. \ref{fig:IV}(b). 
	A low leakage current density of $<\SI{0.09}{\micro\ampere\per\milli\meter\squared}$ under reverse bias is observed, suggesting near-ideal diode behavior. The emission from the neutral exciton is observed when the diode is operated at around $\SI{0.24}{\volt}$ in forward bias, where the current density is lower than $\SI{0.16}{\micro\ampere\per\milli\meter\squared}$. Such current densities are sufficiently low to achieve a stable and low-noise biasing of the quantum dots, that does not add significant spectral diffusion.
	
	The sacrificial layer thickness ($\SI{1370}{\nano\meter}$) of the sample deviates from the optimal thickness of $\SI{1150}{\nano\meter}$\cite{zhou2018}, which reduces the grating efficiency to $\approx 10\%$. 
	The impedance mismatch leads to back reflections from the grating couplers into the waveguide as discussed in the main text.
	We can readily observe the effect of this impedance mismatch from the laser transmission spectrum (c.f. Fig. \ref{fig:IV}(c)) of a rectangular cross-section nanobeam waveguide (width = $\SI{300}{\nano\meter}$; length = $\SI{55}{\micro\meter}$) terminated on each end with a grating coupler.
	The back reflections from the grating coupler results in the formation of a weak cavity across the waveguide that result in the fringes with a free spectral range of $\SI{1.43}{\nano\meter}$. We estimate the reflectivity of the gratings from the fringe amplitude $A$ to be $r^2=(1-A)/(1+A)\approx0.3$ for both exciton and biexciton emission.
	
	\subsection{Glide-plane waveguide characterization}\label{sec:glide}
	The glide-plane symmetry of the GPW features strong in-plane circular polarization in regions of high Bloch mode amplitude. 
	As a consequence, efficient ($\beta\rightarrow1$) and highly directional coupling is abundant in a GPW, i.e. the waveguides are engineered to have chiral light-matter coupling of high Purcell enhancements in large regions within every photonic crystal lattice unit cell \cite{Sollner2015}. \\
	
	The TE-like forward-propagating Bloch mode's electric field $\mathbf{e}_{\mathbf{k},n}(\mathbf{r})$ of mode index $n$, at position $\mathbf{r}$, and wavevector $\mathbf{k}$ is given by $\mathbf{e}_{\mathbf{k},n}(\mathbf{r})=e_{\mathbf{k},n,x}(\mathbf{r})\mathbf{\hat{e}}_x+ e_{\mathbf{k},n,y}(\mathbf{r})\mathbf{\hat{e}}_y$, where $\mathbf{\hat{e}}_x$ and $\mathbf{\hat{e}}_y$ are the cartesian unit vectors. The modes satisfy the Bloch mode normalization
	$\int  \mathrm{d}^3\mathbf{r} \  \norm {\mathbf{e}_{\mathbf{k},n} \left( \mathbf{r} \right)}^2 \epsilon \left( \mathbf{r} \right) = 1$,
	where $\epsilon \left( \mathbf{r} \right)$ is the dielectric permeability.
	The Bloch mode can also be expressed by $e_{\mathbf{k},n,x}(\mathbf{r}) = |e_{\mathbf{k},n,x}(\mathbf{r})|e^{i\phi_{\mathbf{k},n,x}}$ and $e_{\mathbf{k},n,y}(\mathbf{r}) = |e_{\mathbf{k},n,y}(\mathbf{r})|e^{i\phi_{\mathbf{k},n,y}}$, where $\phi_{\mathbf{k},n,x}$ and $\phi_{\mathbf{k},n,y}$ are the phases of the field component along $x$ and $y$ axes, as defined in Fig. 1(a) in the main text.
	Assuming time-reversal symmetry the backward-propagating mode is given by $\mathbf{e}_{\mathbf{-k},n}(\mathbf{r}) = \mathbf{e}^*_{\mathbf{k},n}(\mathbf{r})$.
	
\begin{figure}
		\centering
		\includegraphics[]{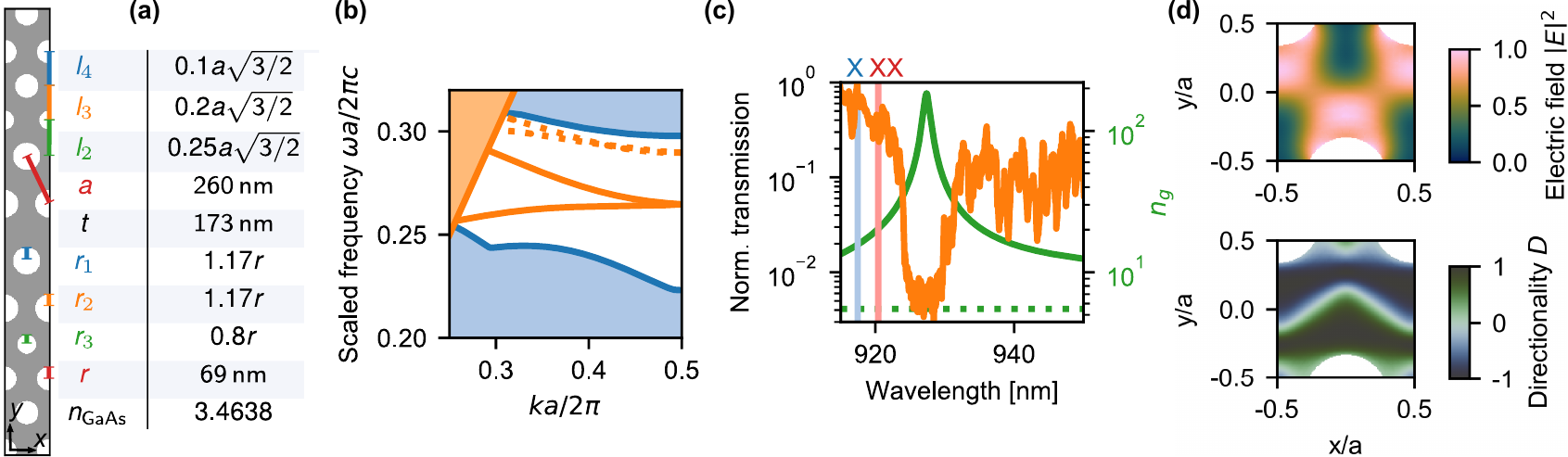}
		\caption{\label{fig:BandDia}\textbf{GPW characterization}. \textbf{(a)} Table of simulation parameters. Here, $t$ is the membrane thickness. Geometry of a unit cell is displayed on the side. The waveguide is formed in the center by repetition along $x$ axis as in Fig. 1(a).  \textbf{(b)} Band diagram of GPW from numerical simulation using COMSOL and the parameters in (a). The geometry is optimized according to Ref. \onlinecite{Mahmoodian2017} in order to obtain single mode waveguide bands. The waveguide modes are displayed with solid lines. The light cone is shaded in orange, while bulk modes are shaded blue. Higher order modes are shown with dotted lines. \textbf{(c)} Identification of the two pass-band regions of the device obtained by scanning a narrow-band laser in a transmission measurement. The emission wavelengths of the $X$ and $XX$ are marked.  Green curves are the simulated group index $n_g$ for the GPW (solid line) and a nanobeam waveguide (dotted line). \textbf{(d)} Top: Local electric field amplitude calculated using COMSOL, the parameters from (a) and $n_{g,X}=14.3$ estimated from the band diagram and the spectral location from the band edge in (b) and (c). Bottom: Local directionality $D_{\mathbf{k},n}\left(\mathbf{r}\right)$ as defined in the main text.}
	\end{figure}

	We calculate the band diagrams $\omega_{\mathbf{k},n}$ and the corresponding Bloch modes $\mathbf{e}_{\mathbf{k},n}(\mathbf{r})$ of the GPW using commercially available finite element software (COMSOL Multiphysics). 
	The model consists of a single supercell with periodic boundary conditions (PBC) along the waveguide direction ($\mathbf{\hat{e}_x}$) and domains of perfectly matched impedance layers (PML) to confine the model. 
	A layer of a perfectly magnetic conductor (PMC) in the vertical symmetry plane of the waveguide reduces the model's numerical complexity, while obeying the TE-like Bloch mode profile.
	The parameters used for the simulation are summarized as a table in Fig. \ref{fig:BandDia}(a). 
	The membrane thickness $t$ is determined during the epitaxial growth while the photonic crystal lattice parameters are defined during the electron beam lithography step in the nanofabrication routine. 
	In the numerical computations, we use the dielectric constant of GaAs at cryogenic temperatures based on the values reported in Ref. \onlinecite{adachi1985gaas}. 
	The resulting band diagram is shown in Fig. \ref{fig:BandDia}(b), where the two fundamental waveguide modes are plotted as solid orange lines. 
	The blue shaded area in the band diagram represents the bulk mode and the orange shaded area corresponds to the light cone
	The bands plotted as dotted lines are the higher order waveguide modes that we do not investigate.
	Further, we determine the group index for of the waveguide modes as ${n}_{g}(\omega_{\mathbf{k},n}) = c / \abs{\nabla_k\omega_{\mathbf{k},n}}$ using the Hellmann-Feynmann theorem \cite{joannopoulos2008molding}: 
	\begin{equation}
	{n}_{g}(\omega_{\mathbf{k},n}) = \frac{2c ( U_{\mathbf{e}_{\mathbf{k},n}} + U_{\mathbf{h}_{\mathbf{k},n}} ) }{\abs{ \int_{\mathcal{V}_s} \mathrm{d}^3\mathbf{r} \ \mathrm{Re}(\mathbf{e}_{\mathbf{k},n}^*(\mathbf{r}) \times \mathbf{h}_{\mathbf{k},n}(\mathbf{r} ))}},
	\end{equation}
	where  $\mathcal{V}_s$ is the total super cell volume, $\mathbf{h}_{\mathbf{k},n}(\mathbf{r})$ is the Bloch mode's magnetic field, and $U_{\mathbf{e}_{\mathbf{k},n}}$ ($U_{\mathbf{h}_{\mathbf{k},n}}$) is the time averaged electric (magnetic) field energy of the Bloch mode.
	
	We characterize the fabricated device through a spectrally-resolved transmission measurement, where a narrow bandwidth continuously tunable laser is coupled into the waveguide at grating coupler $A$ and the light transmitted through the waveguide is collected, coupled off-chip by grating $B$, and measured using a photodiode.
	The measured laser transmission spectrum of a GPW normalized to that of a nanobeam waveguide is shown in Fig. \ref{fig:BandDia}(c) along with the calculated group index $n_g$ (a portion of this scan is also shown in Fig. 2(b) in the main text).
	
	We identify the spectral center of the transmission gap corresponding to the frequency with the highest group index $n_g$. 
	An increase in $n_g$ leads to a loss in transmission due to in-plane optical scattering that scales quadratically with the group index, i.e. $T \propto n_g^2$ \cite{hughes2005prl,Patterson2009prb}. 
	We neglect the dependency on the Bloch mode profile and its dispersion as their contribution is negligible within the spectral width of the transmission gap\cite{NilsSim}. 
	We identify the wavevectors $\mathbf{k}_X$ and $\mathbf{k}_{XX}$ of the exciton and biexciton by their spectral location relative to the frequency with highest $n_g$ (identified in Fig. \ref{fig:BandDia}(c)) and determine the group indices for $X$ and $XX$ as $n_{g,X} = 14.3$ and $n_{g,XX} = 16.1$. 
	The corresponding electric field norm for $n_{g,X}$ is shown in Fig. \ref{fig:BandDia}(d). 
	For the design parameters of the GPW, we observed a $\SI{58}{\nano\meter}$ blue shift of the frequency  with maximum $n_g$ relative to the experimental data, which is consistent with previous results\cite{Pregnolato2019}.\\
	
	Chiral coupling quantifiers such as the directionality\cite{Mahmoodian2017} $D_{\mathbf{k},n}(\mathbf{r})$ and the local phase $\Phi_{\mathbf{k},n}(\mathbf{r})$ can be extracted from the Bloch modes.
	The phase $\Phi_{\mathbf{k},n}(\mathbf{r}) = \phi_{\mathbf{k},n,x}(\mathbf{r}) - \phi_{\mathbf{k},n,y}(\mathbf{r})$ is the relative phase between the electric field components, while the directionality is defined as
	\begin{align}
	\nonumber D_{\mathbf{k},n}\left(\mathbf{r}\right) = &  (\lvert \mathbf{e}_{\mathbf{k},n}\left(\mathbf{r}\right)\cdot\mathbf{\hat{e}_{\sigma^+}}\rvert^2-\lvert \mathbf{e}_{\mathbf{-k},n} \left(\mathbf{r}\right)\cdot \mathbf{\hat{e}_{\sigma^+}}\rvert^2)/|\mathbf{e}_{\mathbf{k},n}(\mathbf{r})|^2 \\
	= &2 \sin{\Phi_{\mathbf{k},n}} |e_{\mathbf{k},n,x}(\mathbf{r})||e_{\mathbf{k},n,y}(\mathbf{r})|/(|e_{\mathbf{k},n,x}(\mathbf{r})|^2 + |e_{\mathbf{k},n,y}(\mathbf{r})|^2)
	\end{align}
	where $\mathbf{\hat{e}_{\sigma^+}}= (\mathbf{\hat{e}}_x+i\mathbf{\hat{e}}_y)/\sqrt{2}$ and $\mathbf{\hat{e}_{\sigma^-}}= (\mathbf{\hat{e}}_x-i\mathbf{\hat{e}}_y)/\sqrt{2}$ are the circularly-polarised unit vectors.
	The calculated directionality for the exciton is shown in the lower panel of Fig. \ref{fig:BandDia}(d). Fig. 1(c) of the main text shows the calculated $\Phi$.\\
	
	\newpage
	\subsection{Quantum dot characterization}\label{sec:SI:QDcharac}
\begin{figure}
		\centering
		\includegraphics[]{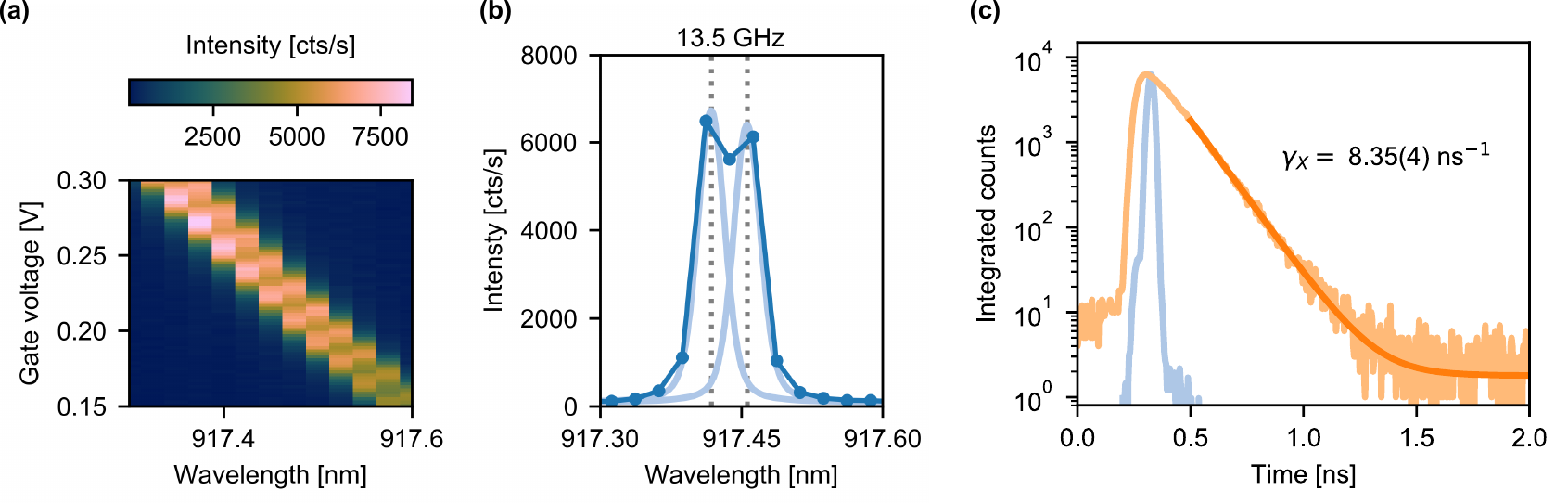}
		\caption{\label{fig:Spectroscopy}\textbf{Quantum dot aboveband spectroscopy} \textbf{(a) }Frequency voltage map of the QD neutral exciton fluorescence under pulsed above band exciation. The two fine structure split levels are visible as two parallel lines, where the  spectrometer resolution of $\sim\SI{25}{\pico\meter}$ limits the visibility. \textbf{(b)} Cut through the frequency voltage map at $\SI{0.248}{\volt}$. The sum of two Voigt profiles is fitted to the spectrum, and plotted with light blue line. The centers of the fitted Voigt profiles are indicated with dotted lines and are split by $\SI{13.5\pm0.5}{\giga\hertz}$. \textbf{(c)} Time resolved fluorescence measurement of the exciton under pulsed non-resonant excitation showing the decay rate of the QD. The decay is fitted to an exponential function convolved with the instrument response function, which is shown in blue. The resulting fit is shown with a solid orange line with the extracted decay rate displayed on the plot.
		}
	\end{figure}

	The QD was identified with above band exciation, using a pulsed diode laser at $\SI{780}{\nano\meter}$. 
	The spectrally-resolved emission from the neutral exciton is seen in Fig. \ref{fig:Spectroscopy}(a) as a function of applied gate voltage where the emission wavelength at the center of the charge plateau is at $\sim\SI{917.45}{\nano\meter}$. 
	Two parallel lines of emission are visible in the map, which correspond to the fine structure split exciton lines. 
	Fig. \ref{fig:Spectroscopy}(b) shows a cut-through of the emission spectrum along $\SI{0.284}{\volt}$. 
	The spectrum is fitted to a sum of two Voigt profiles, which are individually shown as solid light blue lines. 
	The center of the Voigt profiles are highlighted with a dashed line, which allows estimating the fine structure splitting $S= 2 \pi \times \SI{13.5\pm0.5}{\giga\hertz}$.
	Due to the limited spectral resolution of the spectrometer, we observe a small deviation from the value extracted using photon correlation measurements of $2\pi\times \SI{12.78}{\giga\hertz}$.
	
	The QD exciton lifetime is measured in a time-resolved fluorescence measurement. 
	The emission from both the fine-structure-split lines are collected at grating $B$ and detected on an SNSPD, which leads to the histogram displayed in Fig. \ref{fig:Spectroscopy}(c). 
	The histogram is fitted to a single exponential, which yields a decay rate of $\SI{8.35}{\per\nano\second}$. 
	Given the average decay rate of $\SI{3.9\pm0.7}{\nano\second^{-1}}$ for the QDs in the unstructured bulk samples \cite{Chu2020}, the measured exciton lifetime is consistent with the estimated Purcell enhancement of $\approx 2$.
	The single-exponential character of the  decay curve (cf. Fig. \ref{fig:Spectroscopy}(c)) indicates that the two orthogonal linear dipoles have a similar decay rate, and therefore a single exciton decay rate $(\gamma_X)$ suffices in the analysis.
	
	\subsection{Time matching of correlation data}\label{sec:SI:TimeMatch}
\begin{figure}
		\centering
		\includegraphics[]{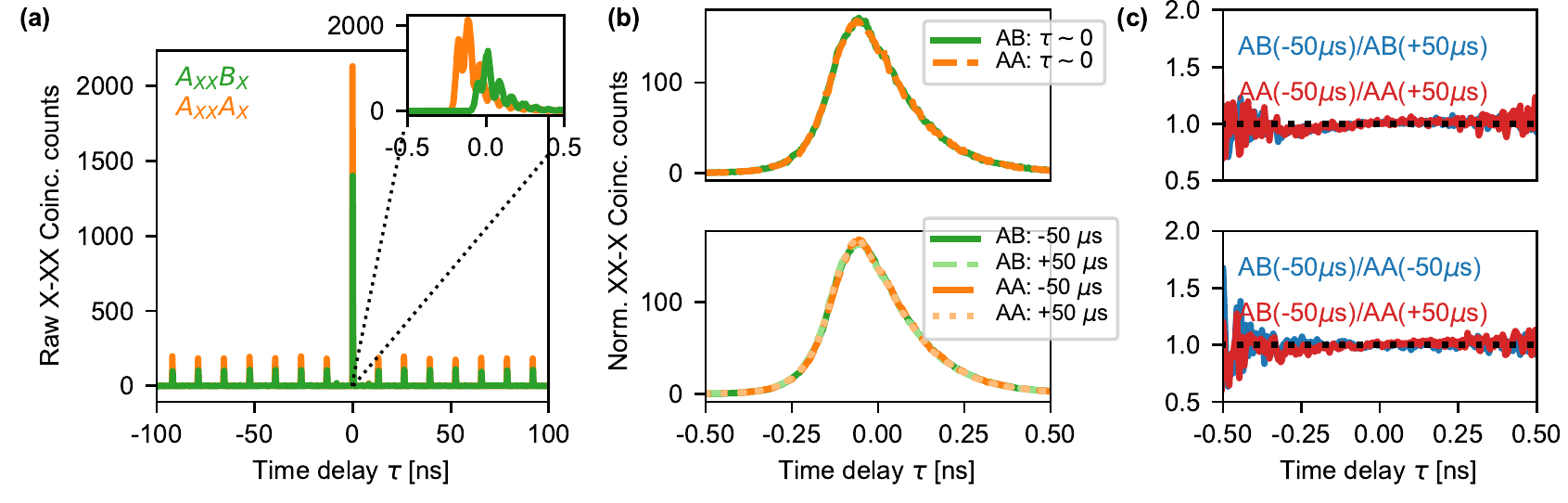}
		\caption{\label{fig:timematch}\textbf{Time matching of different correlation data sets}. \textbf{(a)} Raw data from Fig. 3(a) of the main text, which are offset in time due to different optical path lengths and spectral filtering in the two configurations $A_{XX}B_{X}$ and $A_{XX} A_{X}$. Side peaks where the cascaded bunching is not present are visible at the repetition rate of the laser of $\SI{76.1}{\mega\hertz}$. \textbf{(b) }Side peaks at $\tau \sim \SI{\pm 50}{\micro\second}$ (bottom panel) and $\approx \SI{12}{\nano\second}$ (top panel) on the two configurations after time matching where the high degree of overlap of all measurement configurations is demonstrated with an estimated accuracy better than 40 ppb. \textbf{(c)} Ratio of peaks in each dataset at $\SI{+50}{\micro\second}$ and $\SI{-50}{\micro\second}$ showing accurate estimation of the laser repetition rate and hence the time-zero of the dataset. The increase in the fluctuations about 1 of the ratio at longer time delays is due to the smaller number of detected coincidences.}
	\end{figure}

	The correlation measurements were performed with low-timing jitter ($<\SI{15}{\pico\second}$) SNSPDs (Single Quantum) in order to resolve the fast oscillations arising from the fine-structure splitting. 
	A crucial component in extracting the chirality induced phase shift in correlations is precise time matching of different datasets.
	Even a small variation in the optical path length between the two grating outcouplers results in a timing mismatch between different measured correlation datasets, as seen in inset of Fig. \ref{fig:timematch}(a), which impedes the extraction of the phase shift.
	We collect timestamps of photodetections on the two detectors and process the timestamps to construct coincidence histograms with $\SI{4}{\pico\second}$ bin size over a large time window of $\SI{100}{\micro\second}$.
	The step-by-step procedure implemented for time-matching different correlation histograms is described below with the help of two datasets, $A_{XX}B_{X}$ and $A_{XX}A_{X}$.
	The former correlates exciton emission collected at grating $A$ with that of the biexciton emission collected at grating $B$ while the latter correlates exciton and biexciton emission collected at the same grating ($A$ and $A$).
	\begin{enumerate}
		\item[]{\textbf{Step 1 - Find laser repetition rate:} In order to align the peak at zero time across datasets, we utilize the full timespan, i.e., $\SI{100}{\micro\second}$, to estimate the laser repetition rate $\tau_\mathrm{rep}$ of each dataset using fast fourier transform (FFT). 
			Given the timespan of the dataset, we estimate a FFT resolution of $\SI{7}{\kilo\hertz}$. With a laser repetition rate of $\SI{76.148}{\mega\hertz}$, this FFT resolution corresponds to an imprecision of $\approx\SI{1.3}{\pico\second}$ in the laser repetition period, i.e., $\tau_\mathrm{rep} = 13.1323 \pm 0.0013$ ns.}
		
		\item[]{\textbf{Step 2 - Refine laser repetition rate and find time zero:} From the correlation histogram of the dataset, say $A_{XX}A_{X}$, we select two time-windows of $\tau_\mathrm{rep}$ width (estimated in Step 1), each containing a side peak. 
			These time-windows contain the coincidence peaks at the beginning and the end of the dataset, i.e., at $\tau = \SI{\pm50}{\micro\second}$. 
			By cross-correlating these time-windows separated by $\SI{100}{\micro\second}$, we can improve the precision in estimating $\tau_\mathrm{rep}$. 
			If the cross-correlation function is maximized at zero lag, then the error in $\tau_\mathrm{rep}$ is minimized and $\tau = 0$ is precisely estimated.
			If not, then the lag at which the cross-correlation function maximizes is the amount of adjustment  to be added to $\tau_\mathrm{rep}$.}
		
		\item[]{\textbf{Step 2.1 - Laser repetition rate and time-zero estimation of the second dataset:} The procedure described in Steps 1 and 2 is repeated independently for $A_{XX}B_{X}$ dataset.
			While one could imagine reusing the same repetition rate across all datasets and estimate only time zero point, this naive approach fails across most datasets.
			As the datasets are collected across several days, the repetition rate of the pulsed (Ti:Sapphire) laser showed small variations on the order of a kHz (i.e., $\tau_\mathrm{rep}$ varies by few ps).
			This small variation of $\tau_\mathrm{rep}$ results in errors in time zero estimation of the datasets.
		}
		
		\item[]{\textbf{Step 3 - Align two separate datasets:} 
			We follow the similar procedure as Step 2, but  cross correlate the peak at $ \tau = \SI{-50}{\micro\second}$ from the dataset $A_{XX}B_{X}$ with that of $A_{XX}A_{X}$.
			As discussed in Step 2.1, in order to account for the different laser repetition periods, we rebin the two datasets onto a common time axis.
			Cross-correlating the datasets on this common time axis gives the relative lag in $\tau = 0$ position between them. 
			The lag that maximizes the cross-correlation is used to shift the time-axis of the $A_{XX}B_{X}$ dataset. The resulting aligned datasets $A_{XX}B_{X}$ and $A_{XX}A_{X}$ are shown in Fig. \ref{fig:timematch}(b,c), which shows excellent overlap of the correlation datasets over the complete time span of $\pm \SI{50}{\micro\second}$.
			The performance of the time alignment and $\tau_\mathrm{rep}$ estimation procedure of the two datasets is further verified by dividing the peaks at time delay = $\SI{\pm50}{\micro\second}$ in the dataset $A_{XX} B_{X}$ with their counterpart in the dataset $A_{XX}A_{X}$ as shown in Fig. \ref{fig:timematch}(d), that results in a time-independent ratio that highlights the overlaps of the datasets.}
	\end{enumerate}
	
	Fig. \ref{fig:timematch}(b) shows a pixel-perfect overlap of the individual peaks in datasets $A_{XX}A_{X}$ and $B_{XX}A_{X}$ upon time matching at all time delays $\SI{-50}{\micro\second}\leq\tau\leq\SI{+50}{\micro\second}$.
	As we do not observe any shift between the datasets $A_{XX}A_{X}$ and $A_{XX}B_{X}$ over the entire time span of $\SI{100}{\micro\second}$, we estimate that the misalignment is smaller than the resolution of the time tagger.
	This corresponds to a temporal alignment precision better than $40$ ppb, thereby enabling us to capture the phase shifts of oscillations in the datasets very accurately.
	Crucially, we remove not only the linear time shift between the data but also a nonlinear time shift that arises from imprecision in repetition rate variations.
	While the former is typically easy to handle and requires only a cross-correlation calculation of the entire time span, the latter necessitated the multistep time matching procedure described above.
	Given the oscillation frequency of $\SI{12.78\pm0.03}{\giga\hertz}$ for the QD presented in the main text, it corresponds to a period of $\SI{78}{\pico\second}$.
	As the time matching accuracy is $<\SI{1.3}{\pico\second}$ (worst-case error estimated in Step 1, but improved in Step 2 to $< \SI{1}{\pico\second}$), we conclude that the systematic error from time misalignment in phase shift estimation is minor. 
	
	\newpage
	\subsection{Theoretical model for the state of the biexciton decay in a waveguide}\label{sec:SI:theory}
	
	A path-entangled state can be generated by coupling the emitted photons from a biexciton cascade to a GPW waveguide using chiral light-matter coupling. 
	The fine-structure splitting ($S$), as well as imperfect chiral coupling determine the quality of entanglement. 
	We have modelled the final output state and quantified the quality of the entanglement generated. 
	The details of the calculations will be published elsewhere\cite{bitheory}.
	
\begin{figure}
		\centering
		\includegraphics[]{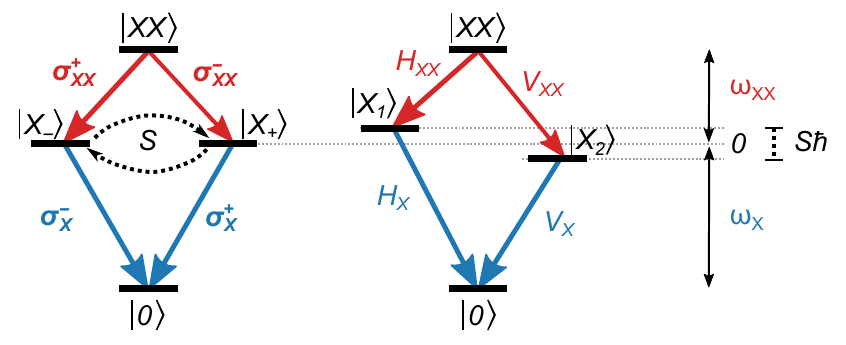}
		\caption{\label{fig:scheme} Energy level structure of the biexciton $\left| XX \right>$ including a fine structure splitting $S$, in different basis representations. The left diagram shows the circular polarization basis of the emitted photons ($\sigma^\pm$), in which the two exciton states $\left| X_{\pm} \right>$ are degenerate in energy. Here the fine-structure splitting induces oscillations between the two exciton states at a rate $\propto 2\pi/S$, as indicated in the figure. Right diagram is in a rotating frame with non-degenerate linearly polarized basis states $\left| X_{1,2} \right>$. The degeneracy is lifted by the fine-structure splitting energy.}
	\end{figure}

	Fig. \ref{fig:scheme} shows the biexciton level structure represented in two different basis descriptions of the exciton states: either two degenerate states $\left| X_{\pm} \right>$ coupled with circularly-polarized transitions $(\sigma^{\pm})$ or non-degenerate states $\left| X_{1,2} \right>$ coupled by linearly-polarized transitions $(H,V)$. 
	In the presence of QD asymmetry, the states $\left| X_{1,2} \right>$ are the eigenstates, while in the former representation the two degenerate eigenstates will oscillate with a characteristic timescale $2\pi/S$. 
	This basis is therefore not stationary in time. 
	To describe the time dependence, we thus use the rotating frame with non-degenerate linear exciton basis $\left| X_{1,2} \right>$ for the complete modelling of the system. 
	
	A wavefunction approach is applied in order to model the experiment in the non-degenerate linear exciton basis by following the theoretical framework put forward in Ref. \onlinecite{Das2019}. 
	The obtained wavefunction components associated with the different decay paths are expressed by Eqs. (2), where we have assumed that the decay rates $\gamma_{XX} = 2 \gamma_{X}$ are the same in horizontal and vertical polarizations, and that they couple equally well to paths $A$ and $B$. This assumption of similar decay rate for the horizontal and vertical is motivated by the fact that  experimentally we only observe a single exponential decay.
	We have further assumed that the chiral phase $\Phi_X=\Phi_{XX}=\Phi$ is identical for biexciton and exciton photons, which is motivated by the comparable $n_g$ at the $X$ and $XX$ emission wavelengths.
	Finally, we have solved the system in the limit in which $S \gg \gamma_X$, i.e. the interference that occurs between the decay of the exciton levels ($\ket{X}_1$ and $\ket{X}_2$) is neglected compared to the effect of the fine-structure splitting when calculating the evolution of their amplitudes.  
	This approximation is well applicable to our system since $S = 2 \pi \times \SI{12.78\pm0.03}{\giga\hertz}$ is an order of magnitude larger than the decay rate $\gamma_X = \SI{8.35\pm0.06}{\per\nano\second}$, i.e. $\Gamma \equiv \gamma_X/(2\pi) = \SI{1.33}{\giga\hertz}$.
	From Eqs. (2) in the main text we calculate the correlation probabilities 
	\begin{align}
	\begin{split}
	P_{A_{XX}A_{X}}\left(\tau\right) &= \abs{\psi_{A_{XX}A_{X}}(\tau)}^2 =  4\gamma_X^2e^{-\gamma_X\tau}\left(1 + \cos\left(S\tau+2\Phi\right)\right) \\
	P_{B_{XX}B_{X}}\left(\tau\right) &= \abs{\psi_{B_{XX}B_X}(\tau)}^2 = 4\gamma_X^2e^{-\gamma_X\tau}\left(1 + \cos\left(S\tau-2\Phi\right)\right)\\ 
	P_{B_{XX}A_{X}}\left(\tau\right) &= P_{A_{XX}B_{XX}}\left(\tau\right) = \abs{\psi_{B_{XX}A_{X}}(\tau)}^2 = \abs{\psi_{A_{XX}B_X}(\tau)}^2 = 4\gamma_X^2e^{-\gamma_X\tau}\left(1 + \cos\left(S\tau\right)\right) \,,
	\end{split}
	\label{eq:prob_appr2}
	\end{align}
	
	which are corresponding to the experimentally recorded correlation data. Here, $\tau>0$ is the time delay between the biexciton and exciton emissions.
	Eqs. (\ref{eq:prob_appr2}) have been used for fitting the time-matched correlation data discussed in \ref{sec:SI:fit} to extract the chiral phase $\Phi$ as shown in Fig. 3 of the main text.
	Note that for perfect chiral coupling $\Phi =\pm\pi/2$, which leads to $P_{A_{XX}A_X}\left(\tau\right)=P_{B_{XX}B_X}\left(\tau\right)$ and out-of-phase oscillations compared to $P_{B_{XX}A_X}\left(\tau\right)=P_{A_{XX}B_X}\left(\tau\right)$. 
	On the contrary, in the non-chiral case ($\Phi = 0,\pm\pi $) all correlations are oscillating in phase.
	As exemplary cases we consider these two extreme situations. 
	Ideal chiral coupling leads to the state
	\begin{align}
	\ket{\psi(\tau)}_{\Phi=\pi/2} = \frac{1}{2}\left( \cos\left(\frac{S \tau}{2}\right)\left( \ket{AB} + \ket{BA} \right) + i \sin\left(\frac{S \tau}{2}\right)\left( \ket{AA} + \ket{BB} \right) \right)\,,
	\end{align}
	which is a maximally path-entangled state. 
	This state oscillates between the Bell states $\ket{\psi^{+}}=1/\sqrt{2}\ket{AB}+\ket{BA}$ and $\ket{\phi^{+}}=1/\sqrt{2}\ket{AA}+\ket{BB}$ with a frequency determined by the fine structure splitting $S$ of the QD. 
	The non-chiral case corresponds to:
	\begin{align}
	\ket{\psi(\tau)}_{\Phi=0,\pi} = \frac{1}{2}\left(  \ket{AB} + \ket{BA} + \ket{AA} + \ket{BB} \right) = \frac{1}{2}\left(  \ket{A} + \ket{B} \right)_X\left( \ket{A} + \ket{B} \right)_{XX}\,,
	\end{align}
	which is a separable state. 
	Non-perfect chiral coupling will lead to a partially-entangled state that can be quantified by the concurrence $C$.
	
	\subsubsection{Entanglement measures} 
	We quantitatively evaluate the degree of the path-entanglement generated in our experiment. 
	For this purpose, we evaluate dependence  various entanglement measures of the state as a function of the chiral phase $\Phi$, as well as the emission and detection time jitters. 
	In particular, the Von Neumann entropy of a two-dimensional bipartite state $\hat{\rho}$ can be obtained as 
	\begin{align}
	H(\hat{\rho}) = - \left( P_1 \log_2\left(P_1\right) + P_2 \log_2\left(P_2\right)\right)\,,
	\end{align}
	where $P_1$ and $P_2$ are the eigenvalues of the reduced density matrix of the system. 
	For compact notation, we omit the $X$ and $XX$ subscripts and follow the convention $AB = A_{XX}B_X$.
	We calculate the density matrix $\hat{\rho} = \ket{\psi}\bra{\psi}$ from the pure state described in Eq. (1) and trace out the biexciton subsystem to get the reduced density matrix
	\begin{align}
	\hat{\rho}_X = \Tr_{XX}\left( \hat{\rho} \right) = \frac{1}{N}\begin{pmatrix}
	\abs{\psi_{AA}}^2 + \abs{\psi_{AB}}^2 & \psi_{AA}\psi^{*}_{BA} + \psi_{AB}\psi^{*}_{BB} \\
	\psi_{AA}^{*}\psi_{BA} + \psi_{AB}^{*}\psi_{BB} & \abs{\psi_{BA}}^2 + \abs{\psi_{BB}}^2
	\label{eq:reduced_rho}
	\end{pmatrix}\,.
	\end{align}
	On the other hand, conservation of the trace under the diagonalization of the reduced density matrix implies that
	\begin{align}
	\Tr\left( \hat{\rho}_x^2 \right) = P_1^2 + P_2^2 = P_1^2 + (1-P_1)^2\,,
	\end{align}
	which allows us to get the eigenvalues directly as a function of the reduced density matrix $\hat{\rho}_X$
	\begin{align}
	P_{1,2} = \frac{1}{2}\left( 1 + \sqrt{2\Tr\left( \hat{\rho}_X^2 \right)-1} \right) = \frac{1}{2}\left( 1 \pm \sqrt{1 - \frac{4}{N^2}\abs{\psi_{AA}\psi_{BB} - \psi_{AB}\psi_{BA}}^2} \right)\,.
	\label{eq:eigenvalue}
	\end{align}
	
\begin{figure}
		\centering
		\includegraphics[]{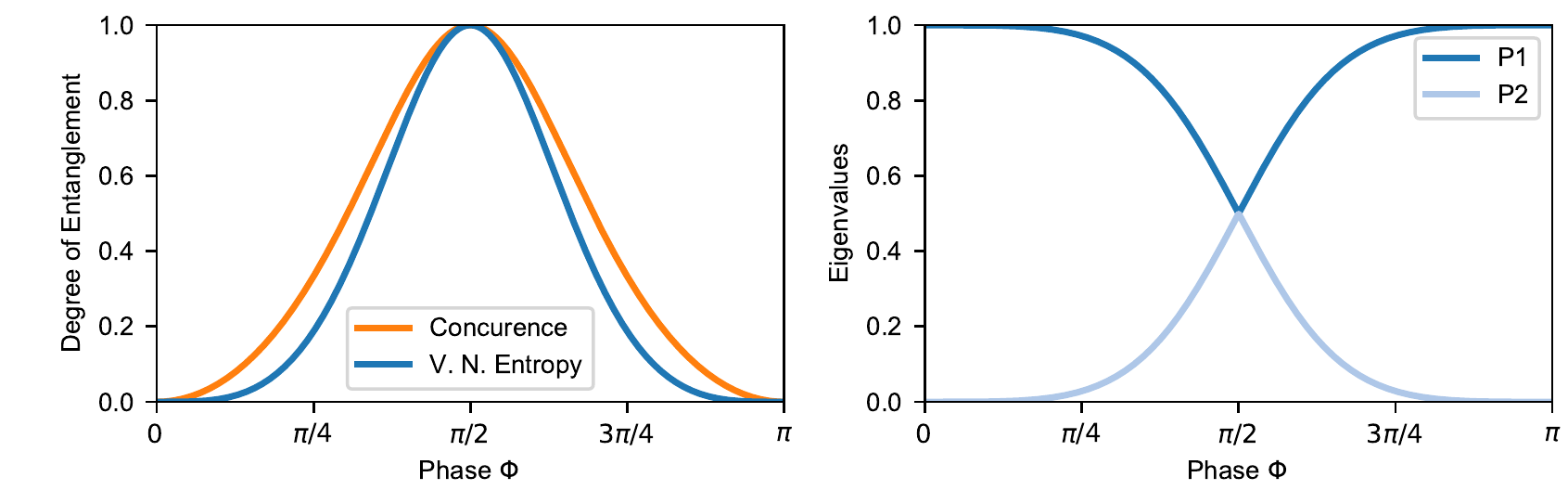}
		\caption{\label{fig:ent1} \textbf{(a)} Von Neumann entropy and concurrence of the path-entangled state as a function of the chiral phase $\Phi$ at $\tau=0$. \textbf{(b)} Corresponding eigenvalues $P_1$ and $P_2$ at $\tau=0$.}
	\end{figure}

	The concurrence $C$ of the state $\hat{\rho}$ is defined as 
	\begin{align}
	C(\hat{\rho})  = \max\{ 0,\lambda_1-\lambda_2-\lambda_3-\lambda_4\}\,,
	\end{align}
	where $\lambda_i$ are the square root of the eigenvalues of $\hat{\rho}\Tilde{\hat{\rho}}$ in descending order and $\Tilde{\hat{\rho}}=(\hat{\sigma}_y\otimes\hat{\sigma}_y)\hat{\rho}^{*}(\hat{\sigma}_y\otimes\hat{\sigma}_y)$ with $\otimes$ representing the direct product and $\hat{\sigma}_i$  being a Pauli matrix, where ${i}\in{x,y,z}$. 
	Substituting $\hat{\rho} = \ket{\psi}\bra{\psi}$ from Eq. (1) and the results from Eqs. (2), the concurrence is
	\begin{align}
	C(\hat{\rho}) = \frac{2}{N}\abs{\psi_{AA}\psi_{BB} - \psi_{AB}\psi_{BA}} = \frac{\sin^2\left(\Phi\right)}{1+\cos\left(S\tau\right)\cos^2\left(\Phi\right)}\,.
	\label{eq:concurrence0}
	\end{align}
	Eq. (\ref{eq:concurrence0}) shows how the fine-structure splitting $S$ and partial chirality $\Phi$ influence the concurrence of the entangled state. A
	QD located at a perfectly chiral position within the waveguide, i.e. $\Phi=\pm\pi/2$, leads to $C=1$, which agrees with the fact that the resultant biphoton state oscillates between the two maximally entangled states as discussed above. 
	Similarly, QD coupled to a non-chiral waveguide ($\Phi = 0,\pm\pi$) leads to $C=0$. 
	Partial chiral coupling results in imperfect entanglement, where $0<C<1$ with a sinusoidally varying concurrence arising from fine-structure-splitting induced oscillations. 
	Both entanglement measures, concurrence $C$ and von Neumann entropy $H(\hat{\rho})$, are plotted in Fig. \ref{fig:ent1}(a) as a function of the chiral phase $\Phi$.
	
	\subsubsection{Timing jitter} 
	Finally we consider the effect of detection time jitter on the entanglement measure, which arises due to the uncertainty in the emission time of the photons, and therefore the detection. 
	We assume a Gaussian distribution of the detection times, motivated by the Gaussian instrument response function of the SNSPDs, with a $\varsigma$ standard deviation centered around a time $\tau'$. 
	The postive-time-averaged ($\tau'>0$) unnormalised matrix element of the density matrix $\hat{\rho}$ is
	\begin{equation}
	\bar{\hat{\rho}}_{ijkl} = \int_{0}^{\infty}d\tau' e^{\frac{-(\tau'-\tau)^2}{2\varsigma^2}}\psi_{ij}(\tau,0,\tau')\psi^{*}_{kl}(\tau,0,\tau'),
	\end{equation}
	where the subscripts represent the collection gratings $\{i,j,k,l\} \in \{A,B\}$. 
	The normalization factor of the state $\bar{N}$ is
	\begin{equation}
	\bar{N} = \int_{0}^{\infty}d\tau' e^{\frac{-(\tau'-\tau)^2}{2\varsigma^2}}N,
	\end{equation}
	where $N$ is the total number of detected photons. 
	The concurrence $C$ is calculated from the density matrix element by numerically computing $\bar{\hat{\rho}}\tilde{\bar{\hat{\rho}}}$ under the same approximation as earlier $S\gg\gamma_X$.
	Figure 3(c,d) of the main text show the expected temporal and spatial dependence of the calculated $C$, respectively.
	As expected, we observe that the ratio $S/\varsigma$ influences the maximum $C$ achievable at a given chiral phase $\Phi$.
	The smaller effect of timing-jitter at shorter times on $C$ observed in Fig. 3(c) is a result of accepting only positive time intervals $\tau'>0$, which effectively constrains the uncertainty at $\tau'\approx0$.

	\subsection{Modelling the experimental data and additional correlation measurements}\label{sec:SI:fit}
	We fit the correlation data in Fig. 3 to the theoretical model presented in \ref{sec:SI:theory} using Eq. (\ref{eq:prob_appr2}). 
	The parameters entering the model are the exciton decay rate $\gamma_X$, the fine structure splitting $S$ and the position of zero time delay $\tau_0$. For the data for the same grating we further introduce that phase $\Phi$, whereas this does not play a role for different gratings.
	The exciton decay rate is extracted separately from Fig. \ref{fig:Spectroscopy}(c). 
	The fine structure splitting estimated in Fig. \ref{fig:Spectroscopy}(b) is used as a starting parameter. 
	Furthermore Eq. (\ref{eq:prob_appr2}) is convolved with a Gaussian function of width $\varsigma$ accounting for the finite detector timing resolution (jitter), and multiplied with an over all amplitude. 
	First we fit the $A$-$B$ correlations, where $\Phi=0$ is fixed. Hereby all other parameters are extracted and used as fixed parameters in the fit of the $A$-$A$ correlations, where the phase is determined. 
	The analysis is performed with Python using lmfit, and the uncertainty in all plots is the $1\varsigma$ confidence interval.
	
\begin{figure}
		\centering
		\includegraphics[]{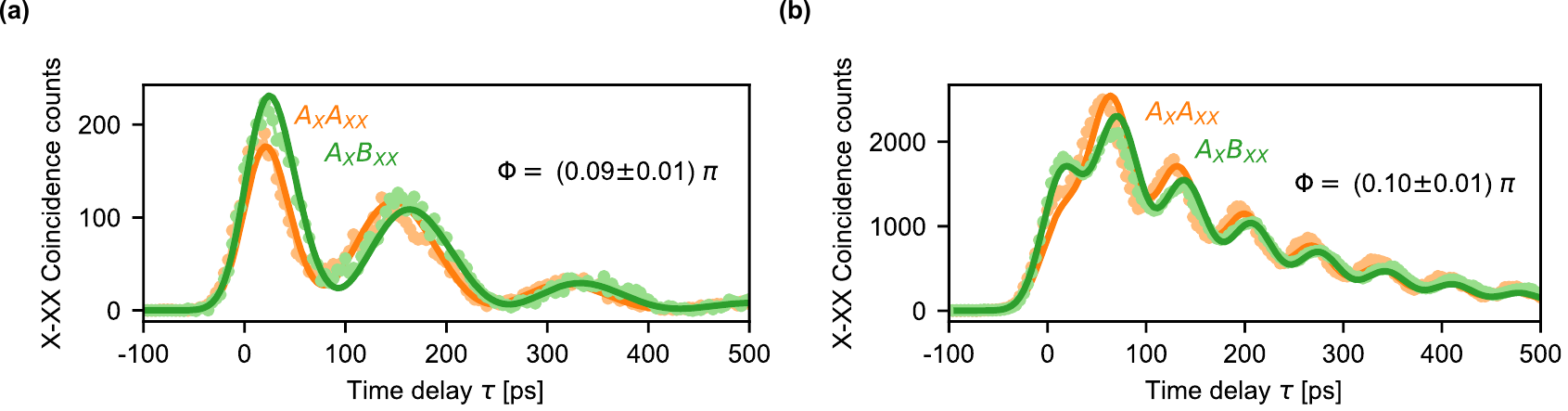}
		\caption{\label{fig:CorrQD23}Correlation measurements of QD2 (a) and QD3 (b) in similar GPWs shown as points and the model as solid lines. The field at position of both QDs exhibit a phase shift as indicated in the figure. Note that the two QDs have different fine structure splittings leading to different oscillation periods.
		}
	\end{figure}

	We found and measured the $XX$-$X$ correlation phase shift of two additional QDs in two other GPW devices with slightly different design parameters, cf. data in Fig. \ref{fig:CorrQD23}. 
	We denote these QD2 and QD3, which have exciton emission wavelengths at $\SI{924.25}{\nano\meter}$ and $\SI{921.55}{\nano\meter}$, respectively. 
	The $X$ emission from QD2 is $\SI{0.25}{\nano\meter}$ from the high $n_g$ region on the high wavelength side, and has a significantly smaller fine structure splitting, leading to slower oscillations. 
	QD3 on the other hand is around $\SI{5}{\nano\meter}$ from the high $n_g$ region on the low wavelength side and has a large $S$, as seen from the faster oscillations.

\begin{figure}
		\centering
		\includegraphics[]{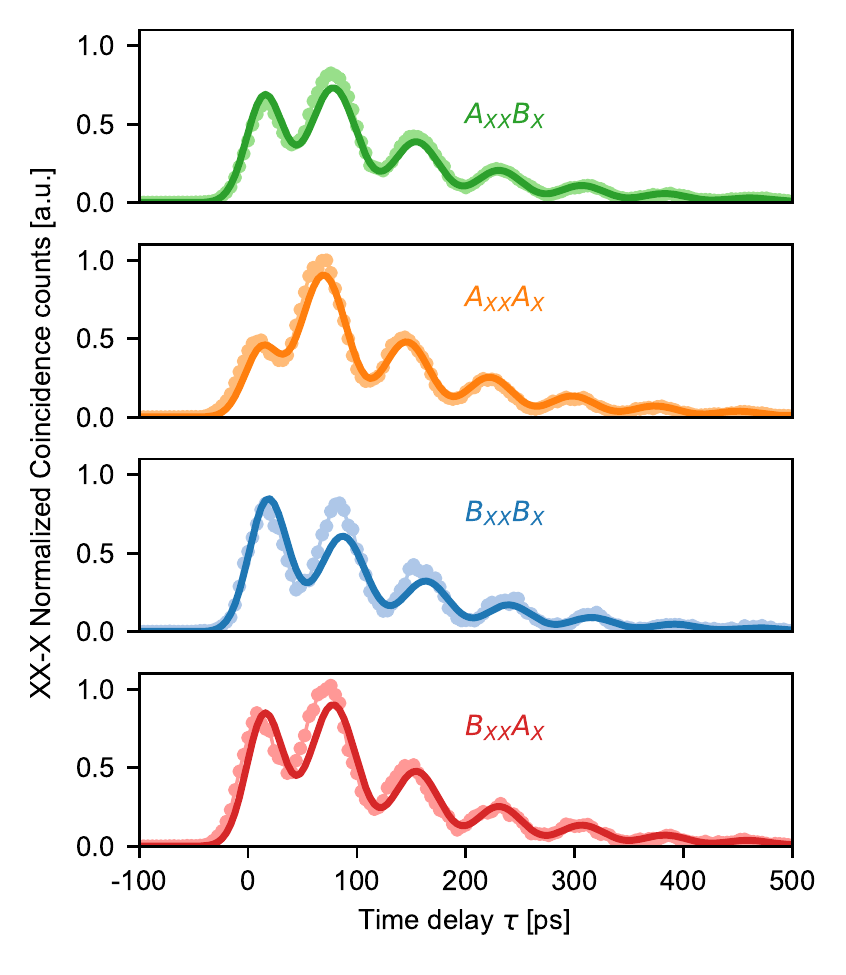}
		\caption{\label{fig:CorrAll}Correlation measurements of QD1 in all four configurations of collection gratings shown as points. All data are time-matched with respect to the data set in the $A$-$B$ configuration. Solid lines show the results from the theoretical model in Eq. (\ref{eq:prob_appr2}) for each configuration, based on the fit result of Fig. 3(a).}
	\end{figure}

	In addition to the correlation data presented in Fig. 3 we performed measurements in the remaining $B_{XX}B_{X}$ and $B_{XX}A_{X}$ configurations as well for that QD. 
	The data are presented in Fig. \ref{fig:CorrAll}. 
	According to the theory (see Eq. (\ref{eq:prob_appr2})) in the same grating configurations ($A_{XX}A_{X}$ and $B_{XX}B_{X}$) we should observe the same phase but with opposite signs. The solid line shows this case superimposed on the data using the fitted phase from Fig. 3(a), where it can be seen that the model describes the data well. 
	
	\subsection{Cause of reduction in the phase}
	We attribute the deviation of the measured $\Phi = 0.12\pi$ from the likely phase values $(\Phi \approx 0.5\pi)$ c.f. Fig 1(b), to the weak back reflections from the grating outcouplers. This is due to a non-optimal design as discussed in \ref{sec:SI:Fab}, where we also estimate the reflectively $r^2$. As discussed in the main text these back reflections will reduce the chirality, due to partial cancellation of the longitudinal component of the electric field. We split the effects into an ideal phase $\Phi_0$ (corresponding to Fig. 1(b) in main text) and a reduced phase $\Phi$, which is the one we can observe in our correlations measurements $\Phi$. 
	
	We model the back reflections from the two grating outcouplers as a cavity with reflectance $r$ in both ends. The propagating-mode component of the electric field is thus modified by added contributions from consequent cavity reflections. The total accumulated component $\tilde{e}_{\mathbf{k},x}$ of the electric field yields
	\begin{equation}
	\tilde{e}_{\mathbf{k},x} = \abs{e_{\mathbf{k},x}}\left( e^{i\phi_x} + re^{i\left(-\phi_x+\theta\right)}+ r^2e^{i\left(\phi_x+\theta+\theta'\right)} + r^3e^{i\left(-\phi_x+2\theta+\theta'\right)} + ...\right) \,,
	\end{equation}
	and similarly for the $y$ component of the field, where $\theta+\theta'\equiv\Theta$ is the total propagation phase in the cavity. Note that here we have opposite phases $\pm\phi_x$ of the forward and backward propagating fields since these modes are the time reverse of each other. Solving the geometric series we obtain
	\begin{equation}
	\tilde{e}_{\mathbf{k},x} = \abs{e_{\mathbf{k},x}}\frac{e^{i\phi_x} + re^{i\left(\theta-\phi_x\right)}}{1-r^2e^{i\left(\theta+\theta'\right)}}\,.
	\end{equation}
	We then calculate the ratio $\tilde{e}_{\mathbf{k},x}/\tilde{e}_{\mathbf{k},y}$ between the electric field components in order to relate the reduced phase $\Phi = \phi_x - \phi_y$ and the initial phase $\Phi_0 = \phi_{0,x} - \phi_{0,y}$ with the reflectance of the cavity. Since we only observe a single exponential decay of the exciton state, c.f. Fig. \ref{fig:Spectroscopy}(c), we constrain the modulus of the x and y components of the field to be the same. This implies that the propagation phase equals the  initial phase $\Phi$. The reflectance can then be calculated to be
	\begin{equation}
	r = \frac{\sin{\left(\frac{1}{2}\left(\Phi_0-\Phi\right)\right)}}{\sin{\left(\frac{1}{2}\left(\Phi_0+\Phi\right)\right)}} \,. \label{eq:refl}
	\end{equation}
	Note that we have set $\tilde{\phi}_y= {\phi}_y = 0$ for simplicity. 
	
	We measure a reflectance of $r=\sqrt{0.3}$ and a phase of $\Phi=0.12\pi$. Applying Eq. \eqref{eq:refl} we obtain that the QD was sitting at $\Phi_0 = 0.37\pi$. Locations with a phase of $0.37\pi$ are likely in the GPW used for the experiment, as seen in the simulation in Fig. 1(b) of the main text. This means that the reduction in phase is can be explained by the measured reflectivity ($30\%$) through Eq. \ref{eq:refl}.
	
	\section*{Acknowledgement}
	We thank Single Quantum for giving us access to the ultra-low timing-jitter superconducting nanowire single-photon detectors required for extracting the correlations. We gratefully acknowledge financial support from Danmarks Grundforskningsfond (DNRF 139, Hy-Q Center for Hybrid Quantum Networks). 
	This project has received funding from the European Union's Horizon 2020 research and innovation programmes under grant agreement No. 824140 (TOCHA, H2020-FETPROACT-01-2018) and under the Marie-Sk\l{}odowska-Curie grant agreement No. 801199.
	A.D.W. and A.L. acknowledge gratefully support of DFG-TRR160,  BMBF - Q.Link.X  16KIS0867, and the DFH/UFA  CDFA-05-06.
	We acknowledge Adam Knorr for initial sample characterization. We thank Matthew Broome for discussions in the early phase of the project.

	\section*{References}

\end{document}